\documentclass[aps,prl,twocolumn,preprintnumbers,superscriptaddress,10pt]{revtex4}

\usepackage{dcolumn,bm}

\usepackage{epsf,amsfonts,amssymb,color}

\usepackage{graphicx}
\usepackage{psfrag}

\newcommand{\be}{\begin{equation}}
\newcommand{\ee}{\end{equation}}
\newcommand{\bea}{\begin{eqnarray}}
\newcommand{\eea}{\end{eqnarray}}
\newcommand{\vev}[1]{{\left< {#1} \right>}}

\newcommand{\eqn}[1]{(\ref{#1})}

\newcommand{\mt}[1]{\textrm{\tiny #1}}
\def\nc {N_\mt{c}}

\newcommand{\pa}{\partial}

\newcommand{\sac}{\, , \qquad}

\newcommand{\uh}{u_\mt{H}}

\newcommand{\n}{n_\mt{D7}}
\def\nc {N_\mt{c}}

\newcommand{\gym}{g_\mt{YM}}

\newcommand{\cf}{{\cal F}}
\newcommand{\cb}{{\cal B}}
\newcommand{\ch}{{\cal H}}

\newcommand{\g}{\gamma}

\begin{document}
\onecolumngrid

\preprint{ICCUB-11-142}
\preprint{MAD-TH-10-06}

\title{The anisotropic ${\cal N}=4$ super Yang-Mills plasma and its instabilities}
\author{David Mateos}
\affiliation{Instituci\'o Catalana de Recerca i Estudis Avan\c cats (ICREA), 
Barcelona, Spain}
\affiliation{Departament de F\'\i sica Fonamental \&  Institut de Ci\`encies del Cosmos (ICC), Universitat de Barcelona, Mart\'{\i}  i Franqu\`es 1, E-08028 Barcelona, Spain}
\author{Diego Trancanelli}
\affiliation{Department of Physics, University of California, Santa Barbara, CA 93106, USA} 
\affiliation{Department of Physics, University of Wisconsin, Madison, WI 53706, USA}

\begin{abstract}
\noindent
We present a IIB supergravity solution dual to a spatially anisotropic 
finite-temperature ${\cal N}=4$ super Yang-Mills plasma. The solution is static and completely regular.  The full geometry can be viewed as a renormalization group flow from an ultraviolet  AdS geometry to an infrared Lifshitz-like geometry. The anisotropy can be equivalently understood as resulting from a position-dependent $\theta$-term or from a non-zero number density of dissolved D7-branes. The holographic stress tensor is conserved and anisotropic. The presence of a conformal anomaly plays an important role in the thermodynamics. The phase diagram exhibits homogeneous and inhomogeneous (i.e.~mixed) phases. In some regions the homogeneous phase displays instabilities reminiscent of those of weakly coupled plasmas.  We comment on similarities with QCD at finite baryon density and with the phenomenon of cavitation.
 
\end{abstract}
\maketitle


\noindent
{\bf 1. Introduction.}
The realization that the quark-gluon plasma (QGP)  produced in heavy ion collisions (HIC) is  strongly coupled  \cite{fluid} has provided  motivation for understanding the dynamics of strongly coupled non-Abelian plasmas through the gauge/string duality \cite{duality} (see \cite{review} for a  review of applications to the QGP). The simplest example of the duality is the equivalence between four-dimensional ${\cal N}=4$  $SU(\nc)$ super Yang-Mills (SYM) theory and  IIB string theory on $AdS_5 \times S^5$. Here we extend this example to the case in which the SYM plasma is \emph{spatially} anisotropic.  For accessibility by a broad audience, details will appear elsewhere \cite{long}. Previous holographic studies of anisotropic plasmas include \cite{include,sing}. One important difference with the gravity solution of \cite{sing} is that the latter possesses a naked singularity, whereas  our solution is completely regular. 

Part of our motivation comes from the fact that the QGP created in HIC is anisotropic. An intrinsically anisotropic hydrodynamic description has been proposed to describe the early stage after the collision \cite{ani}. After that stage each little cube of QGP is isotropic in its own rest frame, but even in this phase certain observables may be sensitive to the physics in several adjacent cubes. 

Weakly coupled  plasmas, both Abelian and non-Abelian, are known to  suffer from instabilities in the presence of anisotropies \cite{instab}. It is therefore interesting to understand whether this also happens in strongly coupled anisotropic plasmas. Our gravity solution  exhibits instabilities reminiscent of weak-coupling instabilities. 


At a more theoretical level, motivation is provided by a connection  with the fluid/gravity correspondence \cite{fluidgravity} and the blackfold approach to black hole dynamics \cite{blackfold}, both of which assert that the effective theory describing the long-wavelength dynamics of a black hole horizon is a hydrodynamic theory. Inclusion of conserved $p$-form charges on the gravity side leads to anisotropic hydrodynamics \cite{blackfold4}.

The IIB supergravity solution that we will present is a finite-temperature generalization of that of~\cite{ALT} and: 
({\it i}\,) it is static and anisotropic; 
({\it ii}\,) it possesses a horizon and it is regular on and outside the horizon; 
({\it iii}\,) it obeys $AdS_5 \times S^5$ asymptotic boundary conditions. 
Staticity is required for simplicity, since e.g.~we would like to study the thermodynamics of the system. The presence of a horizon is dual to the existence of a finite-temperature plasma in the gauge theory. Regularity guarantees that calculations are unambiguous and well defined. The  boundary conditions ensure that holography is on its firmest footing and that the solution is solidly embedded in string theory.

As in \cite{ALT}, we deform the SYM theory by a $\theta$-parameter that depends linearly on one of the three spatial coordinates, \mbox{$\theta = 2\pi \n z$}, where $\n$ is a constant with dimensions of energy. In other words, we add to the SYM action a term \mbox{$\delta S \propto \int  \theta(z) \, \mbox{Tr} \, F \wedge F$}. The system we are describing is therefore a static plasma in thermal equilibrium in the presence of an anisotropic external source. Yet, translation invariance is preserved, since integration by parts yields \mbox{$\delta S \propto - \n \int  dz \wedge \mbox{Tr} \left( A \wedge F + \frac{2}{3} A^3 \right)$}. 

The dual gravity description is as follows. Since the $\theta$-parameter is dual to the IIB axion $\chi$ \footnote{This is the supergravity axion and should not be confused with the gauge theory axion.}, we expect that in the  gravity solution this will be of the form $\chi=a z$. It turns out \cite{long} that $a=\lambda \n/4\pi \nc$, where $\lambda=\gym^2 \nc$ is the  't Hooft coupling. Since the axion is magnetically sourced by D7-branes, the solution can be interpreted in terms of a number of D7-branes wrapped on the $S^5$, extending along the $xy$-directions and   distributed along the $z$-direction with density $n_\mt{D7} = d N_\mt{D7}/dz$ \cite{ALT,long}. For this reason we will refer to $a$ and/or $n_\mt{D7}$ as a `charge density'.  Thus in the gravity description it is clear that isotropy is broken by the presence of anisotropic extended objects. Since their full backreaction is incorporated, the D7-branes are completely `dissolved' in the geometry, just like the $\nc$ D3-branes that give rise to $AdS_5 \times S^5$.  Unlike the case of flavour D7-branes \cite{flavour}, the D7-branes considered here do not extend in the radial direction. Consequently, they do not reach the AdS boundary and they do not add new degrees of freedom to the SYM theory.

As in \cite{ALT}, the solution can be viewed as a renormalization group (RG) flow between an AdS geometry in the ultraviolet and a Lifshitz-like geometry in the infrared. At $T=0$ the Lifshitz metric (in string frame) possesses a naked curvature singularity \cite{long}, but this is hidden behind the horizon at $T>0$. 


\noindent
{\bf 2. Solution.} The ten-dimensional solution is a direct product, one of whose factors is an $S^5$ of constant radius $L$ in the Einstein frame. Therefore it can be viewed as a solution of five-dimensional supergravity with cosmological constant $\Lambda = -6/L^2$. Since only the metric $g$, the axion $\chi$, and the dilaton $\phi$ are excited, it suffices to consider the axion-dilaton-gravity action 
\be
S = \frac{1}{2\kappa^2} \int \sqrt{-g} \left( R + 12 -\frac{1}{2} (\partial \phi)^2 
-\frac{1}{2} e^{2\phi} (\partial \chi)^2 \right) + S_\mt{GH},
\label{action}
\ee
where we have set $L=1$ and $S_\mt{GH}$ is the Gibbons-Hawking boundary term. The Einstein-frame metric is
\bea
ds^2 =  \frac{e^{-\frac{1}{2}\phi}}{u^2}
\left( -\cf \cb\, dt^2+dx^2+dy^2+ \ch dz^2 +\frac{ du^2}{\cf}\right),
\label{sol1} 
\eea
and $\chi=a z$.
Isotropy in the $xy$-directions is clearly respected, but not in the $z$-direction  unless $\ch=1$. The axion induces the anisotropy. $\cf$ is a `blackening factor' that  vanishes at the horizon, $u=\uh$. The boundary is at $u=0$. The dilaton only depends on the radial coordinate $u$, as do $\cf$, $\cb$, and $\ch$, which are completely determined in terms of $\phi$. This in turn obeys a third-order ordinary differential equation which we solved numerically \cite{long}. 
The temperature is determined from the requirement that the Euclidean continuation of \eqn{sol1} be regular, and the entropy density from the area of the horizon. These quantities are well defined since the solution is static, i.e.~the dual plasma is in thermal equilibrium (see Section 5).
\begin{figure*}
\psfrag{x}{$\log\frac{a/T}{}$}
\psfrag{y}{$\log\frac{s/N^2_c}{T^3}$}
\begin{center}
\includegraphics[scale=0.82]{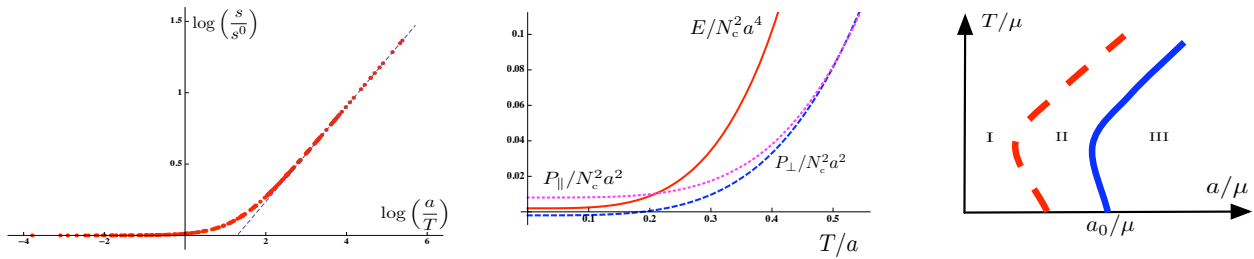}
\caption{(Left) Entropy density as a function of $a/T$. (Center) Energy and pressures as functions of $T/a$ for fixed $a\simeq 2.86$ and $\log \mu=1/2$. (Right) Qualitative phase diagram.
\label{scalings}}
\end{center}
\end{figure*}
Fig.~\ref{scalings}(left) shows the entropy density as a function of $a/T$, normalized by the isotropic value $s^0 (T)= \pi^2 \nc^2 T^3/2$  \cite{correct}.
This provides us with the following check.  We see from the log-log plot that for small $a/T$ the points lie on the horizontal axis, while for large $a/T$ they lie along a line with slope $1/3$. Thus at $T\gg a$ we recover the isotropic result, whereas at $T\ll a$ we recover the Lifschitz scaling $s \propto a^{1/3} T^{8/3}$ found in \cite{ALT}. This interpolating behaviour is expected from the interpretation of the solution as an RG flow.
 
\noindent
{\bf 3. Holographic stress tensor.}
The energy density and the pressures can be obtained from the holographic stress tensor, whose calculation requires the addition of counterterms to \eqn{action}. These can be obtained 
from~\cite{odin,Yiannis} and (in Euclidean signature) take the form
\be
S_\mt{ct}=\frac{1}{\kappa^2} \int d^4x \sqrt{\g}
\left( 3- \frac{1}{8} e^{2\phi} \pa_i\chi\pa^i\chi  \right) 
- \log v  \int d^4x \sqrt{\g} {\cal A}\,,
\label{counterterms}
\ee
where $v$ is the Fefferman-Graham (FG) coordinate, $\gamma$ is the induced metric on a $v=v_0$ surface, and the limit \mbox{$v_0 \to 0$} is understood. ${\cal A}(\g_{ij},\phi,\chi)$ is the conformal anomaly,  which when evaluated on our solution takes the value
${\cal A}(\gamma_{ij},\phi,\chi) = \nc^2 a^4 / 48 \pi^2$.

From the results of \cite{Yiannis} the stress tensor is seen to be diagonal, $\vev{T_{ij}} = \mbox{diag}(E, P_\perp, P_\perp, P_\parallel)$, 
and to obey
\be
\partial^i \vev{T_{ij}}  = 0\sac \vev{T_i^i} = {\cal A} \,,
\ee
thus confirming that  translation invariance is preserved. $P_\perp$ ($P_\parallel$) are the pressures in the $x,y$ ($z$) directions.  As a consequence of the anomaly the transformation of the stress tensor under a rescaling of $a,T$ contains an inhomogeneous piece \cite{kostas,long}, i.e.
\be
\vev{T_{ij} (k a, k T)} = k^4 \, \vev{T_{ij} (a,T)} +  
k^4 \log k \,\, {\cal A} \, h_{ij} \,,
\label{inhom}
\ee
where $h_{ij} = \mbox{diag} \left(1,-1,-1,3 \right)$. In turn, this means that the stress tensor must take the form
\be
\vev{ T_{ij} (a,T)} = a^4  \, t_{ij} \left( a/T \right) 
 + \log \left( a/\mu \right) \,\, {\cal A}  \, h_{ij} \,,
 \label{later}
\ee
where the arbitrary reference scale $\mu$ is a remnant of the renormalization process, much like the subtraction point in Quantum Chromodynamics (QCD). Different choices of $\mu$ are simply different choices of renormalization scheme. 
We emphasize that the presence of this scale implies that \emph{the physics depends on the two dimensionless ratios $T/\mu$ and $a/\mu$, not just on $T/a$.}
Representative plots of the energy and the pressures are shown in 
Fig.~\ref{scalings}(center).

\noindent
{\bf 4. Thermodynamics.} \label{thermo}
As usual, the free energy $F(a,T)=E-Ts=-P_\perp$ is obtained from the on-shell Euclidean action and satisfies 
$( \partial F/\partial T )_a =-s$ \cite{long,blackfold4}. 
Unlike the entropy density, which is scheme-independent, the energy density and the pressures are scheme-dependent (i.e.~depend on $\mu$), but the thermodynamic relations among them are scheme-independent  \cite{long}. We recall that the necessary and sufficient conditions for local thermodynamic stability are 
\be
c_a \equiv  T \left( \partial S/ \partial T \right)_a  > 0\,, \quad
F'' \equiv \left( \partial^2 F/\partial a^2 \right)_T > 0 \,. 
\label{local}
\ee

\noindent
{\bf 5. Phase diagram.}  \label{diag}
Approximate analytic solutions can be found  in the limits $T\gg a, \mu$ and $T\ll a, \mu$, and these suffice to draw the qualitative phase diagram shown in Fig.~\ref{scalings}(right), which we have also verified numerically \cite{long}.
$F''(a,T)$ is negative in Zone I and positive in Zones II and III.  $P_\parallel(a,T)-P^0(T)$ is negative in Zones I and II and positive in Zone III, with \mbox{$P^0(T)= \pi^2\nc^2 T^4/8$} the isotropic pressure. Note that each of the three zones includes points with $T=0$ as well as points with arbitrarily large $a$ and $T$. 

It follows that the homogeneous phase with uniform D7-brane density is 
in stable thermal equilibrium in Zone III. In particular, as can be seen from the positive slope of the continuous red curve in Fig.~\ref{scalings}(center), the specific heat is $c_a >0$ everywhere. Also, the pressures and the energy are all monotonically increasing functions of $T$ at fixed $a$, so the speed of sound in all directions is real and positive. There are no thermal instabilities anywhere in the phase diagram.
 
In contrast, the homogeneous phase is in unstable thermal equilibrium against infinitesimal charge fluctuations in Zone I, where the second condition in \eqn{local} is violated. In Zone II the system is in metastable thermal equilibrium, since it is unstable only against finite charge fluctuations:  the pressure in the $z$-direction is smaller than the pressure of the isotropic phase, and thus bubbles of isotropic phase can form and grow, forcing a compression of the charge in the $z$-direction \footnote{No charge redistribution can occur in the $xy$-directions because the branes extend along those directions.}. In other words, in Zones I and II a carefully prepared homogeneous system 
with initial $(a,T)$ will fall apart into a mixed phase consisting of  high-density anisotropic `droplets' or `filaments' surrounded by isotropic regions \footnote{Generically the dynamical evolution from the initial to the final states will proceed  through far-from-equlibrium states and could be studied e.g.~by  solving the full-fledged time-dependent Einstein's equations.}. The local charge density $a'>a$ will be the same in each of the droplets, and the pressure will exactly equal that of the isotropic phase at the same final temperature, $P_\parallel(a',T')=P^0(T')$. The pair 
$(a',T')$ therefore lies on the continuous blue curve of the phase diagram.  

In Landau's theory of phase transitions the homogeneous phase in Zones I, II and III would be described by a saddle point of the free energy with at least one unstable direction, by a metastable local minimum, and by a stable global minimum, respectively.

\noindent
{\bf 6. Discussion.} \label{discussion}
Our system is in anisotropic thermal equilibrium. This is not surprising, since in the gauge theory isotropy is broken explicitly by an anisotropic external source. The string description makes it clear that the resulting system can be thought of as a fluid with a conserved, isotropy-beaking, two-brane charge (see e.g.~\cite{blackfold4}). 

It is remarkable that our solution is completely regular despite the fact that it incorporates the full backreaction of the D7-branes, whose number scales as  $n_\mt{D7} \sim \nc/\lambda$. Relatedly, we note that the parameter controlling their backreaction, $\lambda n_\mt{D7}/\nc$, is coupling-enhanced as in the case \cite{back} of flavour D7-branes. 

The physics in Zones I and II shares some similarities with that of QCD at low $T$ and finite baryon density \cite{finite}. In that case the pressure of a chirally broken homogeneous phase with density lower than a critical density $n_0$ is negative (except in a tiny region of very small densities). This indicates an instability towards the formation of `droplets' of higher density $n_0$ in which $P=0$ and chiral symmetry is restored, surrounded  by empty space with $n=0$ and $P=0$. In our case, the role of the chirally restored phase is played by the anisotropic phase, the analogue of $n_0$ is $a_0$ (see Fig.~\ref{scalings}(right)), and the `droplets' correspond to the regions of non-zero D7-brane density. These similarities suggest that the transition from the mixed phase to the homogeneous phase may occur via a percolation mechanism, as in some QCD models \cite{percolation} of chiral symmetry restoration \footnote{Note however that in our case percolation takes place effectively in one spatial dimension \cite{long}.}.

The instabilities we have uncovered are reminiscent of instabilities of anisotropic weakly coupled plasmas \cite{instab}. Somewhat pictorially, the main similarity is the tendency to `filamentation', which in weakly coupled plasmas can be understood (very roughly) as the tendency of similarly oriented currents to cluster together. We emphasize though that there are obvious differences. In a weakly coupled plasma the anisotropy is `dynamical' since it arises from the momentum distribution of the particles that compose the plasma. In contrast, in our case the plasma is static and intrinsically anisotropic because of the presence of dissolved extended objects. In any case, we stress that whether a real connection exists between the instabilities studied here and those of  weakly coupled  plasmas is a question beyond the scope of this letter. 

The instabilities of our solution are also reminiscent of  the phenomenon of cavitation, i.e.~the formation of bubbles of vapour in regions of a flowing liquid in which the pressure of the liquid drops below its vapour pressure. Cavitation has been proposed \cite{torri} (see also \cite{also}) as a mechanism that would lead to  fragmentation of the QGP into droplets that would subsequently evaporate, thus providing a new scenario for how hadronization is achieved. In that context the analogue of vapour pressure is the pressure of the vacuum, $P=0$, whereas  in ours it is the pressure of the isotropic phase. As above, however, we emphasize that in the case of \cite{torri} the pressure drop is due to a dynamical effect, namely to the viscosity corrections that result from the expansion of the plasma. In contrast, in our case this is a static effect presumably resulting from the interaction of the extended objects  in the  plasma. 

Note that an instability discovered in \cite{ALT} does not directly apply here, since Lifshitz (instead of AdS) boundary conditions were assumed in \cite{ALT}.



\noindent
{\bf Acknowledgments.}
We are grateful to G.~Grignani and A.~Virmani for collaboration at an earlier stage. We thank A.~Buchel, J.~Casalderrey-Solana, B.~Fiol, J.~Garriga, A.~Hashimoto, E.~Kiritsis, F.~Marchesano, D.~Marolf, G.~Moore, R.~Myers, I.~Papadimitriou, M.~Roberts, K.~Skenderis, and very specially R.~Emparan for comments and discussions. We are particularly  grateful to I.~Papadimitriou for sharing Ref.~\cite{Yiannis} with us prior to publication. 
We are supported by 2009-SGR-168, MEC FPA2010-20807-C02-01, MEC FPA2010-20807-C02-02, CPAN CSD2007-00042 Consolider-Ingenio 2010 (DM) and by PHY04-56556, DE-FG02-91ER40618, DE-FG02-95ER40896 (DT).


\end{document}